\documentclass[aps,prd,twocolumn,showpacs,preprintnumbers,amsmath,amssymb,amsfonts,superscriptaddress,openany,nofootinbib]{revtex4}
\pdfoutput=1 

\usepackage[colorlinks=true,breaklinks=true]{hyperref}
\usepackage[normalem]{ulem}
\usepackage[utf8]{inputenc}
\hypersetup{allcolors=[rgb]{0.0 0.0 0.6},linkcolor=[rgb]{0.75 0.05 0.05}}
\usepackage{amsmath,amssymb}
\usepackage{epsfig}  
\usepackage{graphicx}   
\usepackage{slashed}       
\usepackage{tikz}
\usepackage[compat=1.1.0]{tikz-feynman}
\usepackage{url}
\usepackage{color}
\usepackage{multirow}
\usepackage{comment}
\usepackage{amssymb}
\usepackage{mathtools}
\usepackage{bm}
\usepackage{braket}

\usepackage{multirow}
\usepackage{booktabs}
\usepackage{mathtools}
\usepackage{adjustbox}
\usepackage{array}
\usepackage{caption}
\usepackage{slashed}
\usepackage{tikz,xcolor,hyperref}
\usepackage{comment}

\newcommand{\mr}[1]{\mathrm{#1}}
\usepackage{mathtools}

\newcommand{\yh}[1]{{\color{blue}[Yu: #1]}}
\newcommand{\ag}[1]{{\color{orange}[Anish: #1]}}

\graphicspath{{./fig/}{./}{./figs/}}

\begin{document}

\preprint{DESY-24-198}
\title{Gravitational Wave Mountains:
\it{current-carrying domain walls}
\footnote{This paper is the original version of the article accepted for publication Phys.Rev.D 112 (2025) 11, 115019 with the title ``Gravitational waves created by current-carrying domain walls''}
}

\author{Anish~Ghoshal}
\email{anish.ghoshal@fuw.edu.pl}
\affiliation{
Institute of Theoretical Physics, Faculty of Physics, University of Warsaw, ul. Pasteura 5, 02-093 Warsaw, Poland
}

\author{Yu~Hamada}
\email{yu.hamada@desy.de}
\affiliation{Deutsches Elektronen-Synchrotron DESY, Notkestr. 85, 22607 Hamburg, Germany}
\affiliation{Research and Education Center for Natural Sciences, Keio University, 4-1-1 Hiyoshi, Yokohama, Kanagawa 223-8521, Japan}

\smallskip

\begin{abstract}
Domain wall (DW) networks may have formed in the early Universe following the spontaneous breaking of a discrete symmetry. Notably, several particle physics models predict the existence of current-carrying DWs, which can capture and store particles as zero modes on it. In this study, we demonstrate that gravitational waves (GWs) generated by current-carrying DWs with fermionic zeromodes exhibit a novel feature: an additional peak with a distinct spectral shape in the GW spectrum resembling mountains, arising from metastable topological remnants, which we term ``spherons.'' This distinct signature could be detectable in upcoming GW observatories such as LISA. The results suggest that DW networks in beyond Standard Model scenarios could emit GW signals that are significantly stronger and with greater detectability than previously expected. 
\end{abstract}

\maketitle

\section{Introduction} 
Domain walls (DWs) are typical topological defects that are formed due to cosmological phase transitions in the early Universe when for instance two (nearly) degenerate vacua are present. Just after the formation of the DW network, they evolve in what is known as the scaling regime. During this time the correlation length of the network is approximately the same as that of the Hubble horizon size $L\sim t$ \cite{Press:1989yh,Hindmarsh:1996xv,Avelino:2005kn}. The fraction of the total Universe energy budget stored in the DWs increases linearly with time $\rho_{\rm DW}/\rho_{\rm tot} \propto t$,
which can then easily dominate the total energy density at later stage leading to inconsistency with current cosmological observations;
this is known as the domain-wall problem.
One may avoid this scenario if there is an energy bias $V_{\rm bias}$ between the different vacua. 
Such a bias provides pressure between vacua, which drives the DW network to collapse.\footnote{It could happen that after some time $t_{\rm ann}$ is elapsed, the vacuum energy difference $V_{\rm bias}$ between the two degenerate vacua counterbalances the pressure due to the domain wall surface tension $\sigma$. This may lead DWs toward each other and cause them to annihilate before they can dominate the Universe at a time $t_{\rm dom}$~\cite{Kibble:1976sj,Vilenkin:1981zs,Sikivie:1982qv,Gelmini:1988sf}. Closed DWs at this phase shrink and under specific conditions, may enter within their Schwarzschild radius and form primordial black holes (PBHs)~\cite{Vachaspati:2017hjw,Ferrer:2018uiu,Deng:2016vzb,Deng:2017uwc,Ge:2019ihf,Ge:2023rrq, Gouttenoire:2023gbn,Ferreira:2024eru, Dunsky:2024zdo}, a process known in the literature as ``catastrogenesis''~\cite{Gelmini:2022nim,Gelmini:2023ngs}.}
As reviewed below, the collapsing DW network radiates gravitational waves (GW) with significant amplitude,
which is expected to be observed by GW experiments.
The DW evolution in the early Universe has been widely studied
numerically~\cite{Hiramatsu:2012sc,Hiramatsu:2013qaa,Saikawa:2017hiv,Krajewski:2021jje,Chang:2023rll, Kitajima:2023cek, Ferreira:2023jbu, Kitajima:2023kzu,Dankovsky:2024zvs} as well as analytically~\cite{Press:1989yh,Hindmarsh:1996xv, Avelino:2005kn,Nakayama:2016gxi, Martins:2016ois, Martins:2016lzc, Avelino:2020ubr, Pujolas:2022qvs, Gruber:2024pqh}.

Generically topological defects can have rich internal structure which has the ability to carry some sort of charge without dissipation. 
In the case of cosmic strings~\cite{Witten:1984eb}, a current carried on the strings can prevent the string loop from collapsing,
leading to a stable rotating loop (called vorton~\cite{Davis:1988ij}).
These loops might reach equilibrium configurations due to balance between the string tension and the centrifugal force~\cite{Davis:1988ij,CARTER,Alimi:1990hu} whose classical and quantum stability has been discussed~\cite{Carter:1993wu,Martin:1994jp,Radu:2008pp,Battye:2008mm,Garaud:2013iba,Battye:2021kbd,Battye:2021sji,Ibe:2021ctf,Abe:2022rrh,Harigaya:2024hah}.
Besides the vortons, the cosmological and astrophysical impacts of current-carrying strings have been also discussed~\cite{Ostriker:1986xc,Oliveira:2012nj,Hartmann:2017lno,Brandenberger:2019lfm,Imtiaz:2020igv,Martins:2020jbq,Fukuda:2020kym,Abe:2020ure,Agrawal:2020euj,Theriault:2021mrq,Martins:2021cid,Cyr:2022urs,Rybak:2022sbo,Auclair:2022ylu}.


Similarly, a DW can carry current~\cite{Jackiw:1975fn,Rubakov:1983bb,Kaplan:1992bt,Peter:1995ks,Stojkovic:2000ub,Vachaspati:2006zz},
whose internal degrees of freedom are classified into two cases:
fermionic current carrier arising when the sign of a fermion mass differs on both sides of the DW~\cite{Jackiw:1975fn,Rubakov:1983bb,Kaplan:1992bt,Stojkovic:2000ub}
and bosonic carrier arising when some $U(1)$ symmetry is broken only inside the DW~\cite{Witten:1984eb,Peter:1995ks}.
If DWs are current-carrying, 
one may expect that such a current stabilizes a closed DW  
by balance between the DW tension and the centrifugal force analogously to vortons.
This stabilized object can be long-lived in the Universe, 
and might have some cosmological impact in addition to standard DW networks as we see below.



\section{BSM Model involving DW with currents} 
The concept of the current-carrying DW is quite common.
One simple beyond Standard Model (BSM) example for the fermionic carriers is the DW in two-Higgs doublet model (2HDM)~\cite{Battye:2011jj,Eto:2018hhg,Battye:2020jeu,Battye:2020sxy,Law:2021ing,Sassi:2023cqp,Fu:2024rsm},
where the Higgs potential consisting of two Higgs doublets $H_1,H_2$ has a $\mathbb{Z}_2$ symmetry $H_1 \to H_1$, $H_2 \to -H_2$ with a tiny bias term proportional to $H^\dagger_1 H_2+ \mr{h.c.}$. 
This $\mathbb{Z}_2$ symmetry is spontaneously broken in the vacuum, giving rise to a DW.
Depending on the type of the Yukawa couplings~\cite{Branco:2011iw}, 
the SM fermions have masses whose signs are flipped when getting across the DWs,\footnote{Depending on the parameters, this model can have DWs that break $U(1)$ electromagnetic symmetry by the condensation of the charged bosonic fields~\cite{Battye:2020jeu,Battye:2020sxy,Law:2021ing,Sassi:2023cqp}. In such a case, the bosonic particles also can play roles of the charge/current carriers on the DWs. In this work, however, we do not consider those cases but restrict ourselves to fermionic carriers.}
and hence carry the SM gauge (and baryon/lepton number) current.


Although the 2HDM is quite simple,
the tension of the DW cannot be beyond about $10^{6}\, \mr{GeV}^3$ since it is related to the electroweak scale.
This can be made more general by adding a real SM-singlet scalar $S$ to 2HDM~\cite{Chen:2013jvg,Muhlleitner:2016mzt,Muhlleitner:2017dkd}
whose vacuum expectation value (VEV) is quite general, 
and imposing a $\mathbb{Z}_2$ symmetry $H_1 \to H_1$, $H_2 \to -H_2$, $S \to -S$.\footnote{DW solutions in this model with a different $\mathbb{Z}_2$ symmetry is also possible ~\cite{Sassi:2024cyb}.}
Again, depending on the couplings to the SM fermions, the DW can contain charge/current carriers.
The DW tension is dominantly controlled by the VEV of $S$.

Also one can find a current-carrying DW in $SO(10)$ Grand Unified Theory as shown in Ref.~\cite{Lazarides:1985my}.
In this case, one cannot introduce explicit breaking terms for the discrete symmetry since it is a subgroup of the $SO(10)$ gauge symmetry;
as a result, the decay mechanism of the DWs must rely on nucleation of cosmic string loops~\cite{Kibble:1982dd,Vilenkin:1982ks,Everett:1982nm,Preskill:1992ck,Dunsky:2021tih,Eto:2023gfn,Fu:2024rsm}
or collision with primordial black holes~\cite{Stojkovic:2005zh}.
Nevertheless, we restrict ourselves to cases with bias terms throughout this work.

\section{Domain Walls and GW} It is known that after the production of the DWs,
they form a scaling network~\cite{Press:1989yh,Hindmarsh:1996xv,Avelino:2005kn},
in which the number of the DWs remains about $\mathcal{O}(1)$ per Hubble patch.
One possible way to avoid the DW domination is to introduce in the Lagrangian a tiny bias term breaking the discrete symmetry slightly,
leading to pressure difference $\Delta V$ between different vacua (domains) separated by the DWs.
The DW network collapses when the pressure is comparable to the DW surface energy at $t\simeq t_\mr{ann}$,
satisfying
\begin{align}
 \Delta V \, d_H(t_\mr{ann})^3 \simeq \sigma \, d_H(t_\mr{ann})^2 \quad \therefore t_\mr{ann} \simeq \frac{\sigma}{\Delta V} \, ,
\end{align}
where $d_H$ is the Hubble length and $\sigma$ is the DW tension.
As this must occur before the DW domination era,
which starts at $t_\mr{dom}^{-1} \simeq G \, \sigma$,
we have a necessary condition 
\begin{align}
\Delta V > G \, \sigma^2 \, . \label{eq:V-cond}
\end{align}
with $G$ the Newton constant.

The DW network radiates GW when collapsing.
One may use for a rough estimate the quadrupole formula of the GW emission rate,
from which one can get the radiation rate $\dot{E}_{GW}$ as
\begin{align}
    \dot{E}_{GW} \simeq - G \sigma^2 R(t)^2 \, , \label{eq:GW-quad}
\end{align}
where $R(t)$ is the typical radius of the shrinking DW.
Here we have used that the typical oscillation frequency of the DW is given as $1/R(t)$.
This formula holds for point-like object observed at infinity.
Nevertheless, this gives a nice approximation of the GW amplitude in most cases~\cite{Hiramatsu:2013qaa}.

The radiated GW from DWs might be observed as stochastic GW background at the present Universe.
It is convenient to consider the GW energy spectrum defined as
\begin{align}
 \Omega_\mr{GW} \equiv \frac{1}{\rho_c}\frac{d \rho_\mr{GW}(f)}{d \log f}
\end{align}
where $\rho_\mr{GW}$ and $\rho_c$ are the GW energy density and critical energy density, respectively.

Since the typical length scale of the collapsing network is given by the Hubble size, $R(t_\mr{ann})\sim t_\mr{ann}$, 
the emitted GW spectrum has a peak around $1/t_\mr{ann}$.
Away from the peak, it is well approximated by power-law tails as $f^{3}$ and $f^{-1}$ in IR and UV regimes, respectively~\cite{Hiramatsu:2013qaa,Kitajima:2023cek}.
Here the IR one is deduced from the causality argument.
After the emission, 
the GW spectrum is red-shifted and observed today as
\begin{align}
 \Omega_\mr{GW,0}(f) = \Omega_\mr{GW, max} 
\begin{cases}
\displaystyle \frac{f^3}{f_\mr{peak}^3} & f\ll f_\mr{peak} \\[4ex]
\displaystyle \frac{f_\mr{peak}}{f} & f\gg f_\mr{peak}
\end{cases}
\end{align}
with 
\begin{align}
 \Omega_\mr{GW, max} = 2 \, \times 10^{-3}\, (G \sigma t_\mr{ann})^2 \left(\frac{g_{*0}}{g_{*}(t_\mr{ann})}\right)^{\frac{1}{3}}
\end{align}
and 
$ f_\mr{peak} = t_\mr{ann}^{-1} a(t_\mr{ann})/a(t_0) $ .
Notice that some results that deviate from the UV spectral index $-1$ have been reported in the literature.
For instance, it is estimated to be $-1.7$ in Ref.~\cite{Ferreira:2023jbu} and $-1.5$ or $-1.3$ depending on initial configurations in Ref.~\cite{Dankovsky:2024zvs}.
While they do not affect the GW peak amplitude significantly, the GW signal in the UV regime may be slightly suppressed.

\section{Current-carrying Domain Walls}
\label{sec:current-carryingDW}
When DWs couple to other particles,
the DWs may gain internal degrees of freedom 
by capturing them.
Let us consider a DW consisting of a real scalar $\phi$ coupled to a single Dirac fermion $\psi$ with a Yukawa coupling $y$.
The simplest and self-contained setup is given by the following Lagrangian:
\begin{align}
    \mathcal{L}=\frac{1}{2}(\partial_\mu \phi)^2 - \lambda \left(\phi^2-v_\phi^2\right)^2 + \bar{\psi}\left[i \slashed{\partial} - y \phi \right]\psi \, , \label{eq:Lagrangian}
\end{align}
where the potential term forces $\phi$ to take the VEV $v_\phi$,
leading to the DW configuration.
We here solve the Dirac equation for $\psi$ in the presence of the DW configuration $\phi_\mr{DW}$ located at $x=0$:
$\phi_\mr{DW} = v_\phi \tanh\left(m_\phi x/2\right)$
with $m_\phi$ being the mass of $\phi$ particle.
Ignoring the $y,z$ and $t$ directions reduces to the equation for the $x$ dependence,\
\begin{align}
 i \gamma^1 \partial_x \psi =  y \phi_\mr{DW}  \psi \, .
\end{align}
Since $i \gamma^1$ is hermitian, 
it is always possible to label the solution $\psi$ in terms of two eigenstates corresponding to $i \gamma^1 = \pm 1$.
For $i \gamma^1 = -1$ space,
one gets a solution of
\begin{align}
\psi(x) \propto \varphi \exp \left(-\int^x_0 dx' \phi_\mr{DW}(x') \right)  \, ,
\end{align}
with $\varphi$ a spinor satisfying $i\gamma^1 \varphi=-\varphi$,
from which one can see that this solution is localized at $x=0$ and decays as $|x|\to \infty$.
The solutions dependent on $y,z$ and $t$ are easily obtained by performing the Lorentz boost in the $y$- or $z$-directions on $\psi$.
Note that these solutions are massless modes,
namely, they behave as massless particles propagating only on the DW ($y$ and $z$ directions).
If $\psi$ has a charge,
which can be either a gauge (e.g., electric) or global (e.g., baryon number) charge,
the trapped mode induces current and charge,
leading to the current-carrying DW.

During the time evolution of the DW network,
the network continuously produces closed DWs by reconnection.
However, it has been observed in Ref.~\cite{Dankovsky:2024zvs} that most closed DWs are produced during the network collapsing, $t\simeq t_\mr{ann}$.
Therefore we focus on the latter case.
The former ones may also give additional contributions to our analysis given below.

After it is produced, the closed DWs shrink due to the DW tension and bias,
and may capture particles from the bulk.
This is a crucial mechanism for the DW to get current/charge.
It is convenient to introduce a yield $Y$ for the would-be trapped particles $\psi$ as
$ Y\equiv n_\psi/s$
with $n_\psi$ and $s$ being the number density of $\psi$ and the entropy density in the Universe, respectively.
Here we assume the capture rate of the particles to be $\mathcal{O}(1)$
and the charge of the trapped particles $\psi$ to be maximally asymmetric, namely, without antiparticles
to avoid pair annihilation on the DW.

The total charge captured by the single closed DW with the typical curvature radius $R(t)$ is roughly estimated as
\begin{align}
    Q(R) \simeq \frac{4 \pi Ys}{3}\left(R_0^3 - R(t)^3\right) \, ,\label{eq:charge}
 \end{align}
 where $R_0$ is the typical curvature radius of the produced closed DW, i.e., $R_0$ is around the network size $R_0\sim t_\mr{ann}$.\footnote{Whether the network of the current-carrying DW exhibits the scaling behavior is not trivial. Although we might have deviation from the scaling, we assume such effects not to be significant for our results below. See also Discussion and conclusions.}
One should note that the total current $J$ of the captured particles is negligible due to the cancellation among the trapped particles.
This is because the trapped particles have random direction of the momentum, 
and hence the average of the angular momentum is $0$.

This capture process gives momentum transfer to the DW per unit area and time,
\begin{align}
F_\psi \sim n_\psi\, m_\psi\, v_\mr{DW}   \label{eq:Fpsi}   \, ,
\end{align}
which gives a frictional force on the DW.
Here $v_\mr{DW}$ denotes the DW velocity.
On the other hand, when the captured particles $\psi$ interact with the thermal plasma, 
it gives an additional frictional force,
as given by
\begin{align}
    F_T\sim \alpha T^2 \frac{Q(R)}{4\pi R(t)^2} v_\mr{DW} \label{eq:FT}
\end{align}
with a model dependent parameter $\alpha<\mathcal{O}(1)$,
which corresponds to square of the coupling constant.
The closed DW dynamics is dominated by friction and in \textit{over-damping regime} when either $F_\psi$ or $F_T$ overcomes the force given by the DW tension, i.e.,
\begin{align}
    \mr{max} \left[F_T, F_\psi\right] > \frac{\sigma}{R(t)} \,.\label{eq:damp-condition}
\end{align}
In such a case, 
the closed DW immediately approaches a sphere which shrinks with non-relativistic velocity.

\section{Formation of metastable object --spheron--} 
\label{sec:form-sph}
In order to simplify the following analysis, 
we here assume that the closed DW is in the over-damping regime \eqref{eq:damp-condition},
so that the DW is a slowly shrinking sphere.
As shrinking,
the charge capture process terminates and $Q$ becomes constant at some radius $R(t)$
because the capture is not favored energetically any more. 
(See Appendix~\ref{sup:energy}.)
The trapped fermions are localized around the sphere surface
and spread on the DW surface moving with speed of light, leading to uniform distribution.
This time scale is expected to be faster than DW motion in the over-damping regime.
The motion of each trapped fermion in the direction along the sphere is labeled by the orbital angular momentum $j (= 1/2,3/2, \cdots)$,
see Refs.~\cite{Aoki:2022cwg,Aoki:2022aez,Aoki:2023lqp} and Appendix~\ref{sup:DiracEq}.
The $z$-component of the angular momentum $j_z$ can be taken as
$j_z= -j, \, -j +1, \, \cdots, \,  j$,
which means that there are $2j+1$ degenerated states for each $j$.
A state with the angular momentum $j$ has an energy $ E \simeq (j+\frac{1}{2})/R$
as long as $E\ll m_\psi$.

The trapped fermions should exhibit the Fermi degeneracy with the conserved $Q$.
Denoting the highest energy level of the trapped  particles as $j_\mr{max}$,
(i.e., the Fermi energy is $\sim j_\mr{max}/R$),
the total occupation number ($\sim Q$) is 
\begin{align}
\sum_{j=\frac{1}{2},\frac{3}{2}, \cdots}^{j_\mr{max}} (2j+1) &= \frac{(2j_\mr{max}+1)(2j_\mr{max}+3)}{4}, \\
\therefore j_\mr{max} & \sim \sqrt{Q} \, .
\end{align}
The summation of the energy over $j$ leads to the total energy of the trapped fermions
\begin{align}
    \sum_{j}^{j_\mr{max}} \sum_{j_z=-j}^j E(j)  \sim \frac{j_\mr{max}^3 }{R} \sim \frac{Q^\frac{3}{2}}{R} \, ,
\end{align}
and contributes to the DW sphere energy $M_\mr{DW}$ as
\begin{align}
    M_\mathrm{DW} &= 4 \pi \sigma R(t)^2 + \frac{Q^\frac{3}{2}}{R(t)} + \frac{4 \pi}{3} R(t)^3 \Delta V  \label{eq:MDW} .
\end{align}
Clearly one can see that \eqref{eq:MDW} can have a minimum as a function of $R$,
which means that the trapped charge may prevent them from shrinking due to the centrifugal force of the particles,
corresponding to the second term in \eqref{eq:MDW}.
Then the stabilized radius, denoted by  $R_\mr{sph}$, is determined by the condition
\begin{align}
\left.\frac{d}{d R}M_\mathrm{DW}\right|_{R=R_\mr{sph}} =0  \, .\label{eq:MDW-min}
\end{align}
At $R=R_\mr{sph}$, the DW tension and the centrifugal force of the charge are balanced,
forming a metastable spherical object,\footnote{They would not be spherically symmetric if they had significant total current. Nevertheless, as stated above, they cannot get significant current from the bulk particles. Thus the spherically symmetric configuration is energetically favored.}
which we dub \textit{spheron}.
Such an object is similar to vortons~\cite{Davis:1988ij,Martins:1998gb}, which are made of charge/current-carrying string loop.
Note that the $Q$-dependence of $M_\mr{DW}$ is different from that for vorton energy ($\propto Q^2$)~\cite{Davis:1988ij}.

As stated above, the spherons are assumed to be produced during the network collapse, $t\simeq t_\mr{ann}$ or equivalently $T\simeq T_\mr{ann}(\simeq \sqrt{M_\mr{pl}/t_\mr{ann}})$.
Then one may get
\begin{align}
R_\mr{sph} & \sim \frac{\sqrt{Q}}{(8\pi \sigma)^\frac{1}{3}}\sim \frac{R_0 \sqrt{Y M_\mr{pl} T_\mr{ann}}}{\sigma^\frac{1}{3}} \, , \label{eq:R_sph}
\end{align}
where we have assumed $R_0\gg R_\mr{sph}$ and ignored the bias pressure $\Delta V$ in $M_\mr{DW}$ as it is sub-dominant.

The formation of the spherons can be confirmed by a simple numerical simulation, as shown in Appendix~\ref{sec:simulation}.
This implies that the spheron formation is robust as long as the closed DW has spherical symmetry due to the over-damping regime.
In addition, the formation time scale is controlled by the friction and is typically faster than the cosmological time scale $t_\mr{ann}$.

In order to have stable spherons,
there are a few conditions:
$\psi$ cannot be relativistic in the bulk to be captured by the DW.
Given this, in order to have enough number density of $\psi$,
they should be decoupled from the thermal equilibrium like dark matter.
Furthermore, there is a necessary condition for the trapped mode not to escape into the bulk.
Using the expression of $Q(R_\mr{sph})$ \eqref{eq:charge} and $R_\mr{sph}$~\eqref{eq:R_sph},
the highest energy of the trapped particle (the Fermi energy) is given as
\begin{align}
 E_\mr{max} \sim \frac{j_\mr{max}}{R_\mr{sph}} 
\sim(8\pi \sigma)^{\frac{1}{3}} \, ,
\end{align}
which must be smaller than the bulk mass $m_\psi$
to prevent from escaping into the bulk.
Notice that 2HDM does not contain sufficiently heavy fermions and requires some extension in the matter sector.
For other conditions, see Appendix~\ref{sup:energy}.

\section{Decay of spheron --charge leakage--}
While spherons are classically stable after formed,
they can decay through either quantum decay of charge-carrier particles like the case of vortons~\cite{Fukuda:2020kym,Ibe:2021ctf,Abe:2022rrh,Harigaya:2024hah}
or fission by non-perturbative tunneling like $Q$-balls~\cite{Coleman:1985ki,Lee:1991ax}.
In this section, we focus on the former case: the decay of the trapped fermions $\psi$.
One should be able to calculate this lifetime $t_\mr{dec}$ in principle once the model is fixed.
Nevertheless, 
we keep it as general since it is highly model-dependent and beyond the scope of this work.
The goal of this section is to estimate the GW energy radiated during this process.

Let us assume that 
the decay rate $\Gamma(E)$ of the trapped fermions $\psi$ has a mild dependence on their energy $E$.
The decay takes place at $t\sim t_\mr{dec}$,
and makes a spheron loose its charge,
leading to the time evolution of the charge as $Q(t)=Q(t_\mr{dec}) \exp(-(t-t_\mr{dec})/t_\mr{dec})$,
which is regarded as a charge leakage process from the spheron.

This decay also gives the spheron the angular momentum because of the conservation law.
(Note that the spheron before the decay does not have total angular momentum while each decaying fermion has the angular momentum $j$.)
For simplicity, we consider the $z$-component of the total angular momentum of the decaying fermions, denoted by $J_z$.
As the decay happens randomly among all the trapped fermions,
the decaying fermions are regarded as random samples taken from a population consisting of the initial trapped fermions.
Thus an expectation value of the sample mean for $J_z$ is $\braket{J_z}=0$
while its variance,
denoted by $\sqrt{\braket{J_z^2}-\braket{J_z}^2}=\sqrt{\braket{J_z^2}}$,
is non-zero, 
as calculated based on a mathematical formula between the variance of a sample mean and a population variance,
\begin{align}
    \sqrt{\langle J_z^2\rangle} \sim \sqrt{\frac{Q-\delta Q}{Q-1}\, \frac{\langle J_z^2\rangle_\mr{pop}}{\delta Q}}\, ,
\end{align}
where $\delta Q$ is the number of the decaying fermions (size of the sample) 
and $\langle J_z^2\rangle_\mr{pop}$, the square of the population variance, is the expectation value of $j_z^2$ among all the trapped fermions, given by 
\begin{align}
 \langle J_z^2\rangle_\mr{pop} \equiv \frac{1}{Q}\sum_{j}^{j_\mr{max}}\sum_{j_z=-j}^j j_z^2 \sim j_\mr{max}^2\,.
\end{align}
By assuming $\delta Q$ is comparable to the initial charge $Q(t_\mr{dec})$,
we have
\begin{align}
    \sqrt{\langle J_z^2\rangle} \sim \mathcal{O}(1) \, .
\end{align}
Owing to this, the spheron is no longer spherically symmetric,
and the deviation of the radius $\Delta R$ (with the ellipticity being $1-\Delta R /R$)
is estimated as $\Delta R \sim \sqrt{\langle J_z^2\rangle} /(\sigma R^2)$,
leading to the quadrupole moment $\sim M_\mr{DW} \Delta R\, R \sim M_\mr{DW}/(\sigma R)$.

On the other hand,
as the charge decreases,
the spheron starts to shrink from the radius $R_\mr{sph}$.
There is a $t$-dependent equilibrium point at which the DW tension is balanced by $Q(t)$,
given by $R_{eq}(t)\sim \sqrt{Q(t)}/\sigma^{1/3}$ (see Eq.~\eqref{eq:R_sph}).
Since the decay time scale of $Q(t)$ is much slower than the DW oscillation,
the system is sufficiently adiabatic,
i.e., the radius $R(t)$ just follows the equilibrium point without oscillation. 
The time evolution of the radius $R(t)$ (ignoring the deviation $\Delta R$ due to the angular momentum) is well described by the following equation of motion (EOM)
\begin{align}
 \ddot{R}+2 \frac{1-\dot{R}^2}{R} - \frac{Q^{\frac{3}{2}}}{4\pi \sigma R^4}=0 \, ,\label{eq:effectiveEOM}
\end{align}
which is obtained from the relativistic effective Lagrangian of the DW,
\begin{align}
 S=\int dt \left[-\frac{4\pi \sigma R^2}{\sqrt{1-\dot{R}^2}} - \frac{Q^{\frac{3}{2}}}{R}\right] \, ,
\end{align}
consisting of the well-known Nambu-Goto action of the spherical DW~\cite{Ipser:1983db,Widrow:1989fe} and the static energy coming from the trapped fermions (See Eq.~\eqref{eq:MDW}).
We here ignored the bias contribution and assumed $Q$ to change sufficiently slowly.

\begin{figure}[tbp]
 \centering
 \includegraphics[width=0.45\textwidth]{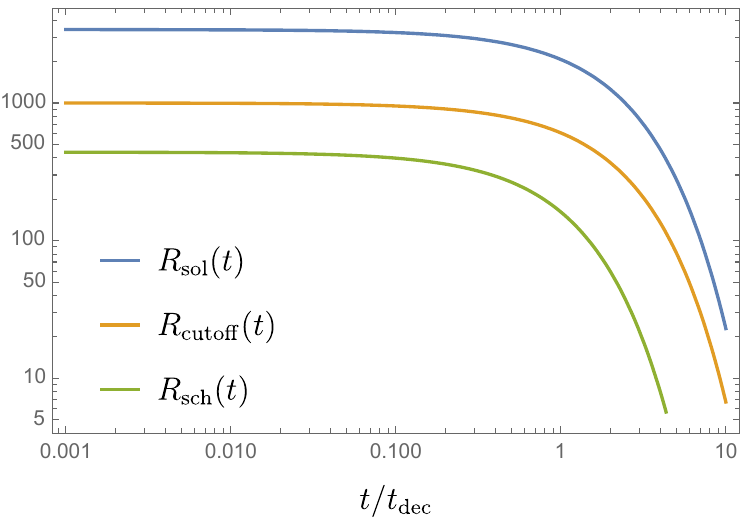}
 \caption{\it
The time evolution of the radius $R_\mr{sol}$ of the decaying spheron in the case of charge leakage, which is obtained by solving Eq.~\eqref{eq:effectiveEOM}.
We here took a dimensionless unit $\sigma^{1/3}=1$ and $G=10^{-6}$.
The cutoff $R_\mr{cutoff}$ is a radius at which the trapped fermions can escape to the bulk, leading to classical instability of the spheron,
which is calculated by $m_\psi=10\, \sigma^{1/3}$.
$R_\mr{sch}$ is the Schwarzchild radius.
}
 \label{fig:decaying_sph}
\end{figure}

The numerical solution of Eq.~\eqref{eq:effectiveEOM} is shown in Fig.~\ref{fig:decaying_sph}.
One can clearly see that the time scale of the evolution of $R(t)$ (blue curve) is comparable to $t_\mr{dec}$.
As references, we also showed the cutoff radius $R_\mr{cutoff}\sim \sqrt{Q(t)}/m_\psi$ (orange),
below which the trapped fermions go to the bulk, leading to the breakdown of the current calculation,
and the Schwarzchild radius $R_\mr{sch}$ (green) calculated from the energy $M_\mr{DW}$ \eqref{eq:MDW} with some benchmark parameters.
From this study, one can estimate the GW radiation power during this process by the quadrupole formula,
given by 
\begin{align}
 \dot{E}_{GW} \sim - G \left(t_\mr{dec}^{-3} M_\mr{DW}(R_\mr{sph})/(\sigma R_\mr{sph})\right)^2\, ,
\end{align}
which is very suppressed compared to the GW from the collapsing network.

\section{Decay of spheron --fission--}
As stated above,
the spherons can decay into smaller spherons with the conserved total charge, which is similar to fission of $Q$-balls.
In this section we show that this is kinematically possible for the spherons following the argument of Q-balls given in Ref.~\cite{Lee:1991ax}
and estimate the GW energy radiated by this process.

We consider a case that the spheron with the charge $Q$ decays into two smaller spherons with the charge $Q-\Delta Q$ and $\Delta Q$, respectively.
The initial energy is calculated by using Eqs.~\eqref{eq:MDW} and \eqref{eq:MDW-min} as
\begin{align}
 M_\mr{DW} = 3(\pi\sigma)^\frac{1}{3} Q + \frac{\Delta V}{6\sigma} Q^\frac{3}{2} \, ,
\end{align}
where we have kept the terms to the linear order of $\Delta V$.
On the other hand, the static energy of the two smaller spherons is given as 
\begin{align}
 M_\mr{DW}' = 3(\pi\sigma)^\frac{1}{3} Q + \frac{\Delta V}{6\sigma}\left((Q-\Delta Q)^\frac{3}{2}+\Delta Q^\frac{3}{2}\right) \, ,
\end{align}
which is smaller than $M_\mr{DW}$ for $0\leq \Delta Q \leq Q$ and becomes minimum at $\Delta Q=Q/2$.
Therefore, this kinematical argument shows that the fission is allowed and is likely to lead to two spherons with half of the initial charge.

\begin{figure}[tbp]
 \centering
 \includegraphics[width=0.45\textwidth]{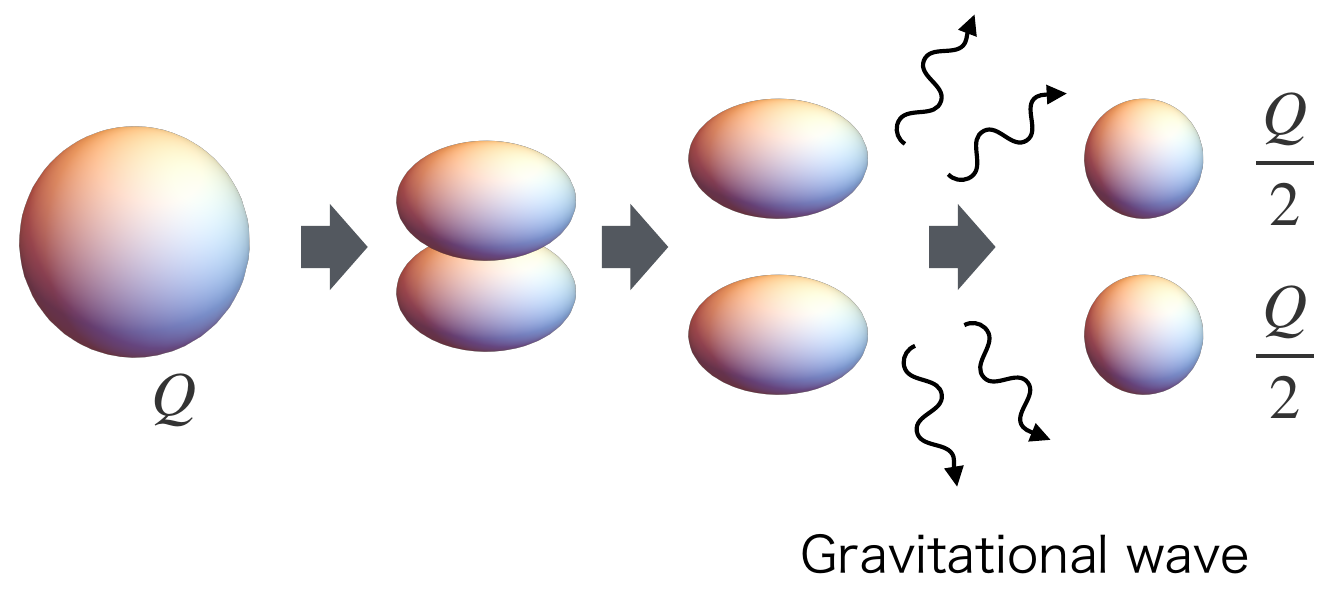}
 \caption{ \it
Schematic illustration of the spheron decaying by the fission.
Right after the fission, the two smaller spherons are significantly deviated from the spherical shape
and can radiate GW to be relaxed.
}
 \label{fig:fission}
\end{figure}

Although the decay rate of the fission is calculable once the model is fixed,
it is quite difficult because it is a non-perturbative process.
We here take the lifetime $t_\mr{dec}$ as a free parameter instead,
and show the potential impact of the spheron decay.
Particularly, we focus on the GW radiation from these decaying spherons.
During the decay process, 
as in the case of tunneling processes described by bounce solutions,
the field configurations ($\phi$ and $\psi$) overcome the energy barrier and reach a configuration with the same energy as the initial energy $M_\mr{DW}$ within the Euclidean time,
after which they ``roll down'' to the energy local minimum $M_\mr{DW}'$ by classical processes with the real time.
This latter stage can radiate GW since the intermediate configuration must be significantly deviated from the spherical shape,
leading to two oscillating excited spherons;
see Fig.~\ref{fig:fission}.

The radiation power from the two spherons is estimated by the quadrupole formula as
\begin{align}
\dot{E}_{GW} \sim -G \left( \frac{M_\mr{DW}}{R_\mr{sph}}\right)^2 \sim - G \left( 12 \pi \sigma R_\mr{sph}\right)^2 \, ,\label{eq:GWpower-fission}
\end{align}
where we have assumed that the two spherons have the mass $M_\mr{DW}/2$ and radius $R_\mr{sph}/\sqrt{2}$.
With this radiation, they will be relaxed to be non-oscillating spheres,
whose total energy is given by $M_\mr{DW}'$.
Thus the radiated GW energy is their difference:
\begin{align}
 E_{GW} \sim -\frac{\Delta V}{6\sigma} Q^\frac{3}{2} \left(1-\frac{1}{\sqrt{2}}\right) \, .
\end{align}
Each smaller spheron in the final state should decay again by a similar process,
whose decay rate is here assumed to be same as the initial one,
and again radiates GW similarly.
This is repeated until the spheron size gets comparable to the DW width $m_\phi^{-1}$,
eventually collapsing to produce massive particles.
Thus total radiated GW energy is given by 
\begin{align}
\left. E_{GW}\right|_\mr{total} \sim -\frac{\Delta V}{6\sigma} Q^\frac{3}{2} \, ,
\end{align}
and its typical frequency is given by the initial size $1/R_\mr{sph}$.

\section{GW spectrum from decaying spherons by fission}
Between the two decay processes of the spherons, the charge leakage and fission,
the former one gives a negligible contribution to the GW radiation.
Thus, we hereafter focus on the fission.
Since each of the decaying spherons radiates GW around $t\sim t_\mr{dec}$,
such a GW is observed as stochastic GW background by observers today. 
Here we calculate the GW spectrum 
as superposition of GW radiated from individual decaying spherons,
ignoring an interference among the multiple spherons.
One should note that this is dominantly radiated at $t=t_\mr{dec}$ with the typical frequency $\sim 1/R_\mr{sph}$,
which results in a different peak frequency from that of the conventional DW network.
Thus we get the GW spectrum as
\begin{widetext}
\begin{align}
 \Omega_\mr{GW,sph} \simeq & \,\frac{8\pi G}{3 H_0^2} \frac{\Delta V}{6\sigma} Q(R_\mr{sph})^\frac{3}{2} \, \frac{1}{t_\mr{ann}^3} 
\left(\frac{a(t_\mr{ann})}{a(t_\mr{dec})}\right)^3
\left(\frac{a(t_\mr{dec})}{a(t_0)}\right)^4
\left(\frac{g_{*0}}{g_{*}(t_\mr{dec})}\right)^{\frac{1}{3}} 
 \times
\begin{cases}
\displaystyle \left(\frac{f}{f_\mr{peak,sph}}\right)^3 & f \ll f_\mr{peak,sph} \\[4ex]
\displaystyle \left(\frac{f_\mr{peak,sph}}{f}\right)^2 & f\gg  f_\mr{peak,sph}
\end{cases} \label{eq:GW-sph}
\end{align}
\end{widetext}
with
\begin{align}
  f_\mr{peak,sph} \equiv \frac{1}{R_\mr{sph}} \frac{a(t_\mr{dec})}{a(t_0)}  \, ,
\end{align}
We have assumed that the spherons remain stable until the decay at $t=t_\mr{dec}$,
giving the number density of the spheron $t_\mr{ann}^{-3}\left(a(t_\mr{ann})/a(t_\mr{dec})\right)^3$.
Here we took the UV spectral index as $f^{-2}$ 
because the radiation power in Eq.~\eqref{eq:GWpower-fission} is proportional to $R_\mr{sph}^2$. 
On the other hand, the IR spectral index is taken to be $f^3$ motivated by the causality argument.
Using Eq.~\eqref{eq:R_sph}, one may rewrite the peak frequency and its maximum value. (See Eqs.~\eqref{eq:Omega-analytic} and \eqref{eq:fpeak-analytic} in Appendix~\ref{sup:analytic}.)


\begin{figure*}[tbp]
 \centering
 \includegraphics[width=0.46\textwidth]{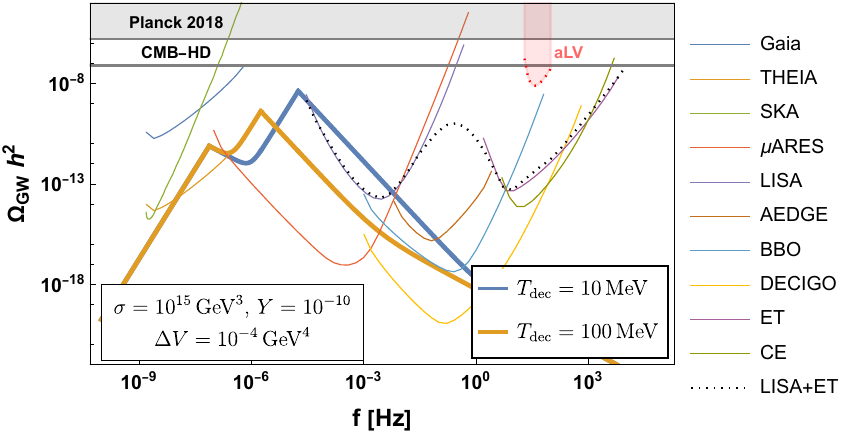}
\hspace{1em} 
 \includegraphics[width=0.46\textwidth]{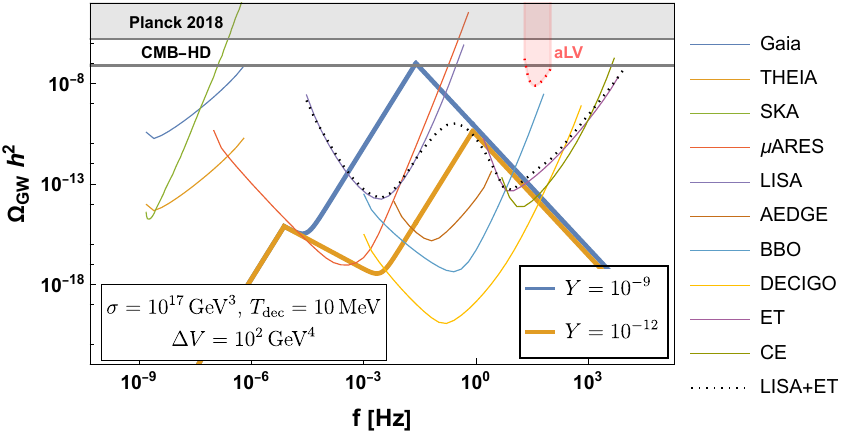}
 \caption{\it
GW spectrum from DW network and spheron. 
In each figure, thick solid curves indicate superposition of those from the conventional DW network (left peaks) and from the spherons which are decaying due to the fission (right peaks).
}
 \label{fig:GW-spectrum-sph}
\end{figure*}

\begin{figure}[tbp]
 \centering
 \includegraphics[width=0.48\textwidth]{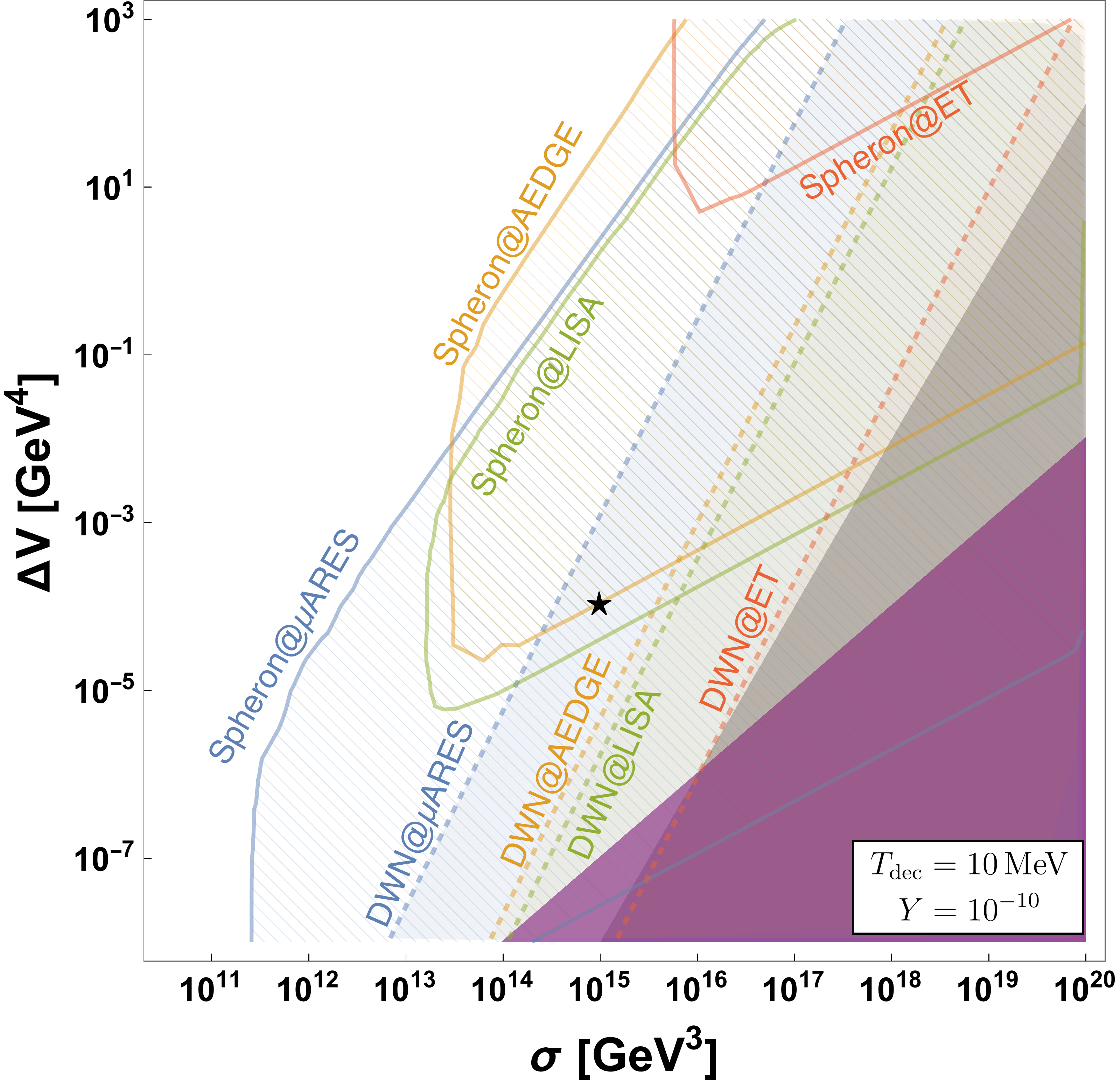}
 \caption{\it
 Parameter space where the two GW signals will be observed with $\mr{SNR} > 10$ (upper-right side of the contours) for different GW experiments.
 We separate the calculation into the signals from the spherons (solid contours) and the conventional DW network (DWN) (dashed lines). 
 The black star indicates the benchmark point corresponding to the blue curve in the left panel in Fig.~\ref{fig:GW-spectrum-sph}.
The gray bottom-right region is excluded due to the DW domination (Eq.~\eqref{eq:V-cond})
while in the purple right-bottom region
the network annihilation is later than the spheron lifetime so that the spheron cannot be formed.
}
 \label{fig:SNR-scan-all}
\end{figure}

Figure~\ref{fig:GW-spectrum-sph} shows the GW spectrum obtained from the superposition of those of the DW network and spherons
with several benchmark cases.
We here assume a step-function-like transition between IR and UV regime for simplicity.
The left and right peaks of thick solid lines correspond to the DW network and spherons, respectively.
Thin solid lines indicate power-law integrated sensitivity curves of future GW observatories:
the Square Kilometer Array (SKA)~\cite{Janssen:2014dka,Weltman:2018zrl},
Gaia and THEIA~\cite{Garcia-Bellido:2021zgu},
LISA~\cite{LISACosmologyWorkingGroup:2022jok, Flauger:2020qyi},
$\mu$ARES~\cite{Sesana:2019vho},
DECIGO~\cite{Kawamura:2020pcg},
AEDGE~\cite{AEDGE:2019nxb},
BBO~\cite{Corbin:2005ny,Harry:2006fi},
Einstein Telescope (ET)~\cite{Punturo:2010zz,Hild:2010id}, 
and
Cosmic Explorer (CE)~\cite{LIGOScientific:2016wof,Reitze:2019iox}. 
We take the threshold signal-to-noise ratio (SNR) to be $\mr{SNR}=1$.
A black dotted curve is obtained by considering LISA and ET operations together, see Ref.~\cite{Marriott-Best:2024anh} for details.
We also show the constraints on $\Delta N_{\rm eff}$ from PLANCK 2018 limits~\cite{Planck:2018vyg}, as well as future reaches of cosmic microwave background (CMB) experiments like CMB-HD~\cite{Sehgal:2019ewc,CMB-HD:2022bsz}.
The red shaded region (aLV) is excluded by the constraint from advanced LIGO-VIRGO~\cite{KAGRA:2021kbb}.
One can see that the GW signal from the spherons is even larger than the conventional one depending on the parameters.
Furthermore, it improves the detectability of GW in the higher-frequency region.

Figure~\ref{fig:SNR-scan-all} shows the parameter space of $\sigma$ and $\Delta V$ in which the SNR exceeds 10.
Solid and dashed contours indicate that individual GW signals from spherons and the conventional DW network are detected by future GW experiments with $\mr{SNR}=10$, respectively.
The gray bottom-right region is excluded due to the DW domination (Eq.~\eqref{eq:V-cond})
while 
the purple region indicates that 
the DW network annihilation time $t_\mr{ann}$ is later than the spheron lifetime so that the spheron cannot be formed.
See Appendix~\ref{sup:snr} for the details of the SNR calculation.
One can find that they have nice complementarity, namely,
GW from spherons can provide significant detectability in parameter space in which that from conventional DW network cannot be detected.
Furthermore, the parameter dependence (especially $\Delta V$ dependence) is quite non-trivial compared to that from DW networks.
This is because the position of the secondary peak (right one) depends on $\Delta V$.

Note that sufficiently long-lived spherons cause the so-called early matter domination (spheron domination),
and their decay injects entropy into the thermal bath and dilutes GW radiated by the DW network~\cite{Gouttenoire:2023pxh}.
The condition to avoid this can be explicitly written down as
\begin{align}
& \frac{M_\mr{DW}(R_\mr{sph})}{t_\mr{ann}^3} 
\left(\frac{a(t_\mr{ann})}{a(t_\mr{dec})}\right)^3 \lesssim T_\mr{dec}^4 \\
&\therefore ~ Y \sigma^\frac{1}{3} \lesssim T_\mr{dec} \, ,
\end{align}
and hence the spheron domination does not happen  in the parameter space presented above.

As stated above, the spheron formation would be robust when the current-carrying DWs feel sufficient friction to exhibit the over-damping regime \eqref{eq:damp-condition},
resulting in the formation efficiency of the order of unity.
This condition \eqref{eq:damp-condition} is realized when $\alpha \gtrsim \mathcal{O}(10^{-8})$ for the parameter space presented in Fig.~\ref{fig:SNR-scan-all}.
On the other hand,
even when the condition is not met,
the formation efficiency may be non-zero
so that the spherons can still have some cosmological impact,
which will be studied elsewhere.

Any source of energy density in the early Universe via its gravitational potential affects CMB perturbations. Unlike the gravitational effect of cosmic strings in the CMB which is typically small unless the string tension is considerably large (see e.g. \cite{Rybak:2021scp}), the situation is not the same when DWs are present due to their growing abundance. 
Previous studies on the CMB and its spectral distortion effects have been studied in the case of stable DW networks~\cite{Zeldovich:1974uw,Lazanu:2015fua,Sousa:2015cqa}, at least up to during the recombination era~\cite{Ramberg:2022irf}.
In addition, the GW generated by the network while in scaling have been shown to have an impact on the CMB B modes using dedicated numerical simulation. 
Such bounds from the GW spectrum of the DW network which translate into bounds on the wall tension $\sigma$ utilising the fact that CMB polarization is able to detect an existing stochastic GW background \cite{Namikawa:2019tax,Ferreira:2023jbu}. 
Nevertheless, those studies do not apply to our parameter space because the DW network and even the spherons disappear well before the recombination era,
$T_\mr{ann},T_\mr{dec} \gg 1\,\mr{eV}$.
In principle, if the spheron is long-lived enough, 
one should be able to put bounds on the parameter space via the CMB constraints,
which we leave as future studies.

\section{DW interpretation of PTA signal}
\label{sec:pta}

\begin{figure}[tbp]
 \centering
 \includegraphics[width=0.45\textwidth]{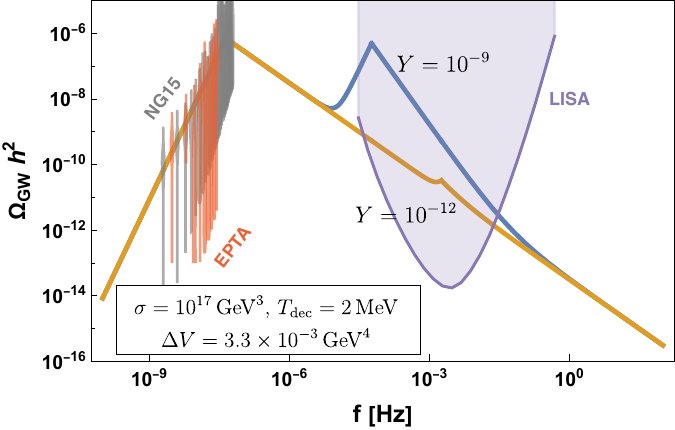}
 \caption{ \it
GW spectrum from DW network and spheron. 
We took the parameters $\sigma$ and $\Delta V$ such that the GW spectrum from the network fits recent PTA signals (NANOGrav: gray, EPTA: red).
If one fixes $T_\mr{dec}=2\, \mr{MeV}$, the additional peak from spherons can lie within LISA sensitivity range (purple) for $10^{-12}\lesssim Y \lesssim 10^{-9}$.
}
 \label{fig:GW-spectrum-PTA}
\end{figure}

We here investigate the DW network interpretation of the stochastic GW background recently reported in pulsar timing array (PTA) collaborations~\cite{NANOGrav:2023gor,EPTA:2023fyk,Xu:2023wog,Reardon:2023gzh}.
In order for the GW spectrum from the DW network to fit the PTA signals, 
we should take appropriate $\sigma$ and $\Delta V$,
e.g., $\sigma=10^{17}\, \mr{GeV}^3$ and $\Delta V=3.3\times 10^{-3}\, \mr{GeV}^4$.

If these DWs are current-carrying, we may have an additional peak from spherons as stated above,
so that we should be able to predict the position and height of the additional peak.
In particular, 
one finds the peak within the sensitivity range of LISA 
when one takes $T_\mr{dec}=2\, \mr{MeV}$ and $10^{-12}\lesssim Y \lesssim 10^{-9}$, see Fig.~\ref{fig:GW-spectrum-PTA}.
(For $Y\lesssim 10^{-12}$, it is difficult to recognize the additional peak.)
The gray and red bars show the ``violin plots'' for NANOGrav with $15$-year data~\cite{NANOGrav:2023gor} and the European Pulsar Timing Array~\cite{EPTA:2023fyk}, respectively.
The purple shaded region is the power-law integrated sensitivity for LISA, calculated in the same way as Fig.~\ref{fig:GW-spectrum-sph}.

\medskip

\section{Discussion and conclusions} 
%
We showed that GW arising due to current-carrying DW leads to a novel shape in the form of an extra peak.
It is noteworthy that the spectral index of the new contribution is distinct from that of the standard DW network,
which would help to distinguish them in the GW experiments.
%

Finally let us comment on several future perspectives.
We discussed above simple BSM examples involving 2HDM (+ singlet scalar extension). 
While our analysis is model-independent, our prescription for the GW spectrum and the parameter space that be tested can be easily translated into BSM microphysics. 
In addition, as stated in Sec.~\ref{sec:current-carryingDW},
the current-carrying DW can feel more friction than usual~\cite{Huang:1985tt,Vilenkin:2000jqa,Nakayama:2016gxi,Blasi:2022ayo,Hassan:2024bvb}.
It is non-trivial how this affects the evolution of the DW network before collapsing.
Furthermore, the formation of spherons can be studied in more detail,
for example for cases without significant friction on the DWs,
and for cases with a spread in the formation epoch, 
which would qualitatively smear the resulting GW spectrum, potentially broadening the peaks.
A further detailed analysis is necessary to address these points.
It is also interesting to consider the coexistence of the spherons and PBHs formed by the collapsing network.

It would also be useful to compare our setup with other exotic objects such as vortons, Q-balls, and Fermi-balls~\cite{Hong:2020est}. 
In particular, GW emission from vorton decay has not been studied in detail. 
By contrast, the decay of Q-balls and Fermi-balls does not radiate GWs because their spherical symmetry is preserved, which is crucial for PBH formation from Fermi-balls~\cite{Kawana:2021tde}. 
Nevertheless, a sudden transition from the Q-ball-dominated era to the radiation-dominated era can induce second-order GWs via scalar perturbations~\cite{White:2021hwi,Kasuya:2022cko,Kawasaki:2023rfx}
with large anisotropic components~\cite{Yu:2025jgx}.
Our analysis above is therefore complementary to these studies and may help to discriminate GWs sourced by different exotic objects.
Studies on GW anisotropies from spherons would also be helpful in this regard.

In summary, Gravitational Wave Mountains from current-carrying DW are an interesting target for planned GW searches.
Ultimately, our results call for an independent confirmation based on numerical lattice simulations,
which will be done elsewhere.


\section*{Acknowledgments}
The authors thank 
Simone Blasi,
David Dunsky,
Ricardo Z. Ferreira,
Marco Gorghetto,
Yann Gouttenoire,
Yuta Hamada, 
Ryusuke Jinno,
Thomas Konstandin,
Kohta Miura, 
Kyohei Mukaida, 
Wakutaka Nakano,
Sabir Ramazanov,
Mohamed Younes Sassi,
and Tanmay Vachaspati
for useful discussions.
This work is supported by 
the Deutsche Forschungsgemeinschaft under Germany's Excellence Strategy - EXC 2121 Quantum Universe - 390833306.

\appendix

\section{Conditions for spheron stability}
\label{sup:energy}
In order to have sufficient charge/current on DW, 
we have two conditions.
The first condition: $m_\psi$ should not be much smaller than the kinetic energy of $\psi$ in the bulk.
This comes from the fact that the typical scale of the trapping potential that $\psi$ feels is $m_\psi$.
Thus $\psi$ cannot be trapped on the DW but must be transmitted or reflected when kinetic energy of the injecting particle is larger than $m_\psi$.
Owing to this condition, one finds that $\psi$ cannot be relativistic in the thermal bath since relativistic particles have kinetic energy $T$( $> m_\psi$).
Given this, in order to have enough number density (or $Y$) of $\psi$,
they should be decoupled from the thermal equilibrium like dark matter.
One particular possible example is asymmetric dark matter,
in which the dark matter abundance in the bulk is dominated by the asymmetric part,
leading to the natural suppression of pair annihilation of the trapped charge on the DW.
Even without such an asymmetry in the bulk,
it is possible to consider the DW with a significant CP violation,
which captures particles and antiparticles with different probabilities,
resulting in asymmetry only on the DW.
See, e.g., Ref.~\cite{Sassi:2023cqp} for CP-violating DW in 2HDM.
When there is no asymmetry on the DW, 
still one could have enough charges due to statistical fluctuation.

The second condition: the capture process must be energetically favored compared to transmission or reflection.
This is because otherwise the capture process is not favored to happen, leading to termination of the capture process.
When the DW sphere radius varies from $R$ to $R-\Delta R$,
this condition is given as
\begin{align}
& \left| - \frac{Q^\frac{3}{2}}{R^2} + \frac{\partial M_\mr{DW}}{\partial Q} \frac{\partial Q}{\partial R}\right| \Delta R  \leq 4\pi n_\psi m_\psi\,  R^2 \Delta R 
\end{align}
where the lhs indicates the increase of the energy due to the capture while the rhs is the energy of the free particles in the case that the capture does not happen.
For $R$ to be close to $R_0$, 
the two terms in the lhs vanish,
resulting in the inequality being satisfied.
This means that the capture is significant at least just after the DW starts to shrink,
and this stage gives a dominant amount of the captured particles.
This allows us to assume the $\mathcal{O}(1)$ capture rate as the first study.


\section{Solution of Dirac equation}
\label{sup:DiracEq}
We here solve the Dirac equation for trapped modes in the background of a spherically symmetric closed DW,
\begin{align}
    \left[ i \slashed{\partial} - m_\psi \hat\phi(r,t)\right] \psi & = 0 \, , \label{eq:Dirac-4d}
\end{align}
where 
$\hat\phi\equiv \phi/v$ is a dimensionless scalar field
and we assume $\phi$ to be almost static, i.e., negligible $t$-dependence hereafter.
The Dirac equation can be rewritten in terms of the one-particle Hamiltonian $\hat{H}$,
\begin{align}
    i\partial_0 \psi = \hat{H}\psi 
\end{align}
with
\begin{align}
    \hat{H} \equiv \gamma^0 \left[ -i\gamma^i \partial_i + m_\psi \hat\phi(r)\right] \, .
\end{align}
Since $\hat{H}$ is hermitian, 
it has a real eigenvalue $E$.
Using the Dirac representation of the $\gamma$-matrices,
we have
\begin{align}
    \hat{H} = 
    \begin{pmatrix}
        m_\psi \hat\phi(r) & -i \sigma^i \partial_i \\
        -i \sigma^i \partial_i &  -m_\psi \hat\phi(r)
    \end{pmatrix} \, ,
\end{align}
where $\sigma^i$ is the Pauli matrices.
In the limit of $E\ll m_\psi$, 
the eigen functions corresponding to trapped modes are given as~\cite{Aoki:2022aez,Aoki:2022cwg}
\begin{align}
    \psi_{j,j_z} \simeq \frac{c}{r} \exp \left[-m_\psi \int _0 ^r dr' \, \hat\phi(r') \right] \, 
    \begin{pmatrix}
    \chi_{j,j_z} \\
   i \frac{\sigma \cdot x}{r} \chi_{j,j_z}
    \end{pmatrix}, \label{eq:eigenmode}
\end{align}
where $\sigma\cdot x\equiv \sigma^i x^i$,
$c$ is a normalization constant determined by
\begin{align}
1 = \int_0^\infty dr \, 4\pi r^2 \, \psi_{j,j_z}^\dagger \psi_{j,j_z} \, ,
\end{align}
leading to $c\propto m_\psi^{1/2}$,
and $\chi_{j,j_z}$ is an eigen function of total angular momentum $\hat{J}$ and $z$-component of angular momentum $\hat{J}_3$,
\begin{align}
\hat{J}^2\,\chi_{j,j_z} &= j(j+1) \,\chi_{j,j_z} \quad \left(j=\frac{1}{2},\frac{3}{2},\cdots \right)\\
\hat{J}_3\,\chi_{j,j_z} &= j_z\, \chi_{j,j_z} \quad (j_z=-j,\cdots,j) \, .
\end{align}
Although $\chi_{j,j_z}$ depends on spatial angles on $S^2$, we suppress them for notational simplicity.
These equations clearly show that for a given $j$ there is $2j+1$ degeneracy. 
In this limit, the eigenvalue $E$ is explicitly written in terms of the radius of the DW $r_0$ (i.e., $\hat\phi(r_0) = 0$) as
\begin{align}
    E \simeq \frac{j+\frac{1}{2}}{r_0}  \left(1+\mathcal{O}((r_0 m_\psi)^{-1}\right)\, . \label{eq:E-eigen}
\end{align}
We here consider $E$ as a positive value because we are interested in the case in which only particles are captured on the DW instead of anti particles.
The anti particles have eigenvalues $E\simeq-(j+\frac{1}{2})/r_0$ and the same angular momentum as the particles.

Then the general solution of the Dirac equation Eq.~\eqref{eq:Dirac-4d} is given by
\begin{align}
    \psi = \sum_{j,j_z} a_{j,j_z}\, e^{-iEt}\, \psi_{j,j_z}  + \cdots\,, \label{eq:mode-expd}
\end{align}
where $a_{j,j_z}$ is a coefficient of the mode expansion to be considered as an annihilation operator associated with the modes of $j$ and $j_z$ after quantization,
and the second term ``$\cdots$'' denotes non-localized modes (scattering states) and the negative-energy modes,
which we are not interested in.

\section{Numerical simulation of spheron formation}
\label{sec:simulation}

\begin{figure*}[tbp]
 \centering
 \includegraphics[width=0.4\textwidth]{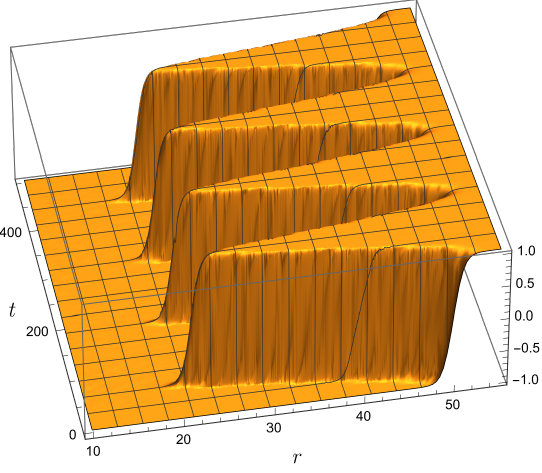} \hspace{4ex}
 \includegraphics[width=0.4\textwidth]{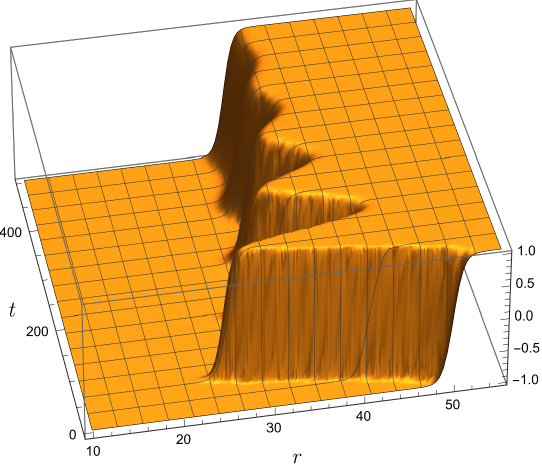} 
 \caption{\it 
3D plots for spheron formation. The time evolution of $\phi(r,t)$ with the fermionic term $\bar \psi \psi$ in the EOM.
 It starts from a large DW sphere with zero velocity.
 The total occupation number $Q$ is fixed as $Q=2022$. 
 (Left): No friction term in the EOM.
 Owing to the energy conservation, the DW sphere oscillates for long time.
 (Right): A friction term $\partial_t\phi(r,t)$ is additionally introduced in the EOM.
 The configuration converges to a stable spheron.
}
 \label{fig:phi-3d}
\end{figure*}

\begin{figure}[tbp]
 \centering
 \includegraphics[width=0.45\textwidth]{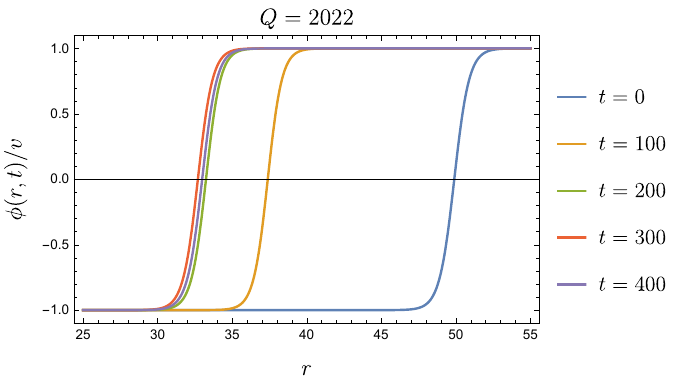} 
 \caption{\it 
 Plots for spheron formation.
 The same simulation as the right panel in Fig.~\ref{fig:phi-3d}.
 The configuration converges to a stable spheron.
}
 \label{fig:phi-2d}
\end{figure}

Here we give a preliminary numerical simulation to explore the formation of spherons.
We solve the time evolution of the scalar field $\phi$, whose EOM is obtained from Eq.~\eqref{eq:Lagrangian}:
\begin{align}
    v^2 \partial_\mu\partial^\mu \hat\phi + \frac{\delta V[\hat\phi]}{\delta \hat\phi} + m_\psi \bar{\psi}\psi  & =0 \, .\label{eq:phi-eom} 
\end{align}
For simplicity, we ignore the bias term and consider a spherically symmetric closed DW,
from which the system becomes essentially $(1+1)$-dimensional problem in $(r,t)$ space.
Our strategy is to treat the fermionic operator $\bar{\psi}\psi$ in the EOM in a ``semi-classical'' way.

At any given time $t$, 
the fermionic Fock space is determined through the mode expansion Eq.~\eqref{eq:mode-expd}, obtained by solving the Dirac equation~\eqref{eq:Dirac-4d}.
The Fock space itself depends on $\phi$ and hence implicitly the time $t$, denoted by $\mathcal{H}[\phi(r,t)]$.

Let us suppose that at $t=0$, 
fermionic particles are trapped on the DW with the Fermi degeneracy and the fermionic state is given by
\begin{align}
    \ket{j_\mr{max};t=0} = \prod_{j}^{j_\mr{max}} \prod_{j_z=-j}^{j} a^\dagger_{j,j_z}\ket{0}_{t=0} \in \mathcal{H}[\phi(r,t=0)] \, ,
\end{align}
where all states with $j\leq j_\mr{max}$ are occupied while the others with $j>j_\mr{max}$ are empty.
This state is the lowest-energy state with a given occupation number $Q=(2j_\mr{max}+1)(2j_\mr{max}+3)/4$
and symmetric under the spatial rotation, resulting in the total angular momentum being zero,
which is self-consistent with the spherically symmetric DW.
The energy gap to the first excited state that also has the vanishing total angular momentum is of order of $j_\mr{max}/r_0$.

As the DW moves, 
the fermionic state $\ket{j_\mr{max};t}$ evolves as well.
This makes it complicated to calculate the time evolution generically.
Nevertheless, it is simplified
if the DW motion is adiabatic and captures no more particles.
The former is equivalent to stating that the time scale of $\phi$ ($\simeq m_\phi$) is much smaller than $j_\mr{max}/r_0$.
Therefore, the Fermi degeneracy is not spoiled and the state at arbitrary time $t$ is given by the corresponding state in the Fock space $\mathcal{H}[\phi(r,t)]$:
\begin{align}
    \ket{j_\mr{max};t} = \prod_{j}^{j_\mr{max}} \prod_{j_z=-j}^{j} a^\dagger_{j,j_z}\ket{0}_t \in \mathcal{H}[\phi(r,t)] \, .
\end{align}

Then,
utilizing the mode expansion Eq.~\eqref{eq:mode-expd} at the time $t$,
the operator $\bar{\psi}\psi$ is evaluated as
\begin{align}
   \bra{j_\mr{max};t} \bar{\psi}\psi \ket{j_\mr{max};t} 
   &= \sum_{j}^{j_\mr{max}}\sum_{j_z}  \psi^{\dagger}_{j,j_z} \gamma^0 \psi_{j,j_z}\, .
\end{align}
Note that this vanishes at the leading order of the $m_\psi \gg E$ limit,
which is shown by substituting Eq.~\eqref{eq:eigenmode} into $\psi_{j,j_z}^\dagger \gamma^0 \psi_{j,j_z}$.
Thus one needs to derive the eigenfunctions at the sub-leading order.
Note that such a contribution is of order of $\mathcal{O}(1/m_\psi)$,
whose $m_\psi$-dependence is canceled with that in the squared normalization constant $c^2$,
resulting in $\psi_{j,j_z}^\dagger \gamma^0 \psi_{j,j_z}\propto \mathcal{O}((m_\psi)^0)$.

Since the calculation for the sub-leading order is quite complicated and beyond the scope of this paper, 
we instead write it down with a general form.
First, the radial part of $\psi_{j,j_z}$ should contain the exponential shape even at the sub-leading order.
Second, by the dimensional analysis, $\psi^{\dagger}_{j,j_z} \gamma^0 \psi_{j,j_z}$ must be proportional to $r_0^{-3}$.
(Note that this is of the order of $(m_\psi)^0$ as stated above.)
Thus, in terms of an undetermined constant $\kappa $, we have
\begin{align}
    \psi^{\dagger}_{j,j_z} \gamma^0 \psi_{j,j_z}\simeq \frac{\kappa}{r_0^3} \exp \left[ -2m_\psi \int _0^r dr' \, \hat\phi(r')\right]  
\end{align}
which leads to 
\begin{align}
    &\bra{j_\mr{max};t} \bar{\psi}\psi \ket{j_\mr{max};t}  \nonumber \\
    &\simeq \frac{\kappa\left(j_\mr{max}+\frac{1}{2}\right)
    \left(j_\mr{max}+\frac{3}{2}\right)}
    {r_0^3} \exp \left[ -2m_\psi \int _0^r dr' \, \hat\phi(r')\right]   \label{eq:psibar-psi}
\end{align}

With Eq.~\eqref{eq:psibar-psi} in the scalar EOM~\eqref{eq:phi-eom},
we perform a numerical simulation in $1+1$ dimensions to explore the spheron formation.
We take an initial configuration to be a sufficiently larger sphere of the DW at $t=0$ and evolve it with the fourth-order Runge-Kutta method.
At initial stages the effect from the fermion term $\bar \psi \psi$ is not significant since it is proportional to $r_0^{-3}$.
As the sphere shrinks, however, it becomes important and gives repulsion.
Eventually the DW sphere bounces back due to the repulsion and expands for a certain time,
and afterwards shrinks back again. 
This oscillation persists forever in this setup because the system is perfectly spherically symmetric and there is no way to lose its energy.
This is shown in the left panel in Fig.~\ref{fig:phi-3d}.
Here we took $\kappa=0.07$, $y=2$, $\lambda=0.5$, and the total occupation number $Q=2022$.
In this dimensionless unit, $v$ is normalized to be $1$.

On the other hand, there must be friction for $\phi$ in reality, such as the Hubble friction and/or friction $F_\psi, F_T$ (see Eqs.~\eqref{eq:Fpsi} and \eqref{eq:FT}, respectively).
This provides dissipation for the dynamics of the DW and hence makes it converge to a static configuration, which is nothing but the spheron.
In order to imitate this, we introduce a frictional term $\gamma \partial_t \phi$ in the EOM~\eqref{eq:phi-eom} with a constant $\gamma$.
Although this $\gamma$ might be quite small in the real evolution in the early Universe, 
we take a relatively large value $\gamma=0.015$ to see its convergence within our simulation.
The result is shown in the right panel in Fig.~\ref{fig:phi-3d} and Fig.~\ref{fig:phi-2d}.
We confirmed that this converged radius (spheron radius) is proportional to $\sqrt{Q}$ in our simulation.
This is consistent with the energetic argument Eq.~\eqref{eq:R_sph}.
Note that $\kappa$ is generally not determined by the above calculation,
and hence the radius depends on $\kappa$ like $R_\mr{sph}\propto \sqrt{\kappa}$.
Thus we here took it such that the converged radius agrees with Eq.~\eqref{eq:R_sph}.

\section{Analytic expression of GW spectrum}
\label{sup:analytic}
Using Eq.~\eqref{eq:R_sph}, one may rewrite the peak frequency and its maximum value of the GW spectrum \eqref{eq:GW-sph} as 
\begin{widetext}
\begin{align}
\left. \Omega_\mr{GW,sph}\right|_\mr{max} \simeq & \,
5\times 10^{-11} \,
\left(\frac{Y}{10^{-12}}\right)^\frac{3}{2}
\left(\frac{10^{18}\, \mr{GeV}^3}{\sigma}\right)^\frac{1}{4}
\left(\frac{\Delta V}{10^{2}\,\mr{GeV}^4}\right)^\frac{1}{4} \left(\frac{T_\mr{dec}}{10\, \mr{MeV}}\right) \, ,\label{eq:Omega-analytic}
\end{align}
\begin{align}
 f_\mr{peak,sph} 
\simeq 
 3\times 10^{-1} \, \mr{Hz} \left(\frac{10^{-12}}{Y}\right)^\frac{1}{2} \left(\frac{\Delta V}{10^{2}\,\mr{GeV}^4}\right)^\frac{3}{4} 
\left(\frac{10^{18}\, \mr{GeV}^3}{\sigma}\right)^{\frac{5}{12}}
  \left(\frac{10\, \mr{MeV}}{T_\mr{dec}}\right) \, ,\label{eq:fpeak-analytic}
\end{align}
\end{widetext}
from which one can see that the parameter dependence is much different from the DW without current/charge.
Thus there is an optimized $\Delta V$ leading to the maximum SNR for each GW experiment. 
Even for large $\Delta V$,
higher frequency GWs with larger $\Delta V$ can be detected in ultrahigh frequency GW detectors in the future, see Ref.\cite{Aggarwal:2020olq} for a review.
Note that this expression holds only for $R_0 \gg R_\mr{sph}$,
which is rewritten as $\sigma^{7/3} \gg Y^2 M_\mr{pl}^3 \,\Delta V$.
Without this condition, it is difficult to write down its analytical expression.

\section{Signal-to-noise ratio (SNR)}
\label{sup:snr}

\begin{figure}[tbp]
 \centering
 \includegraphics[width=0.45\textwidth]{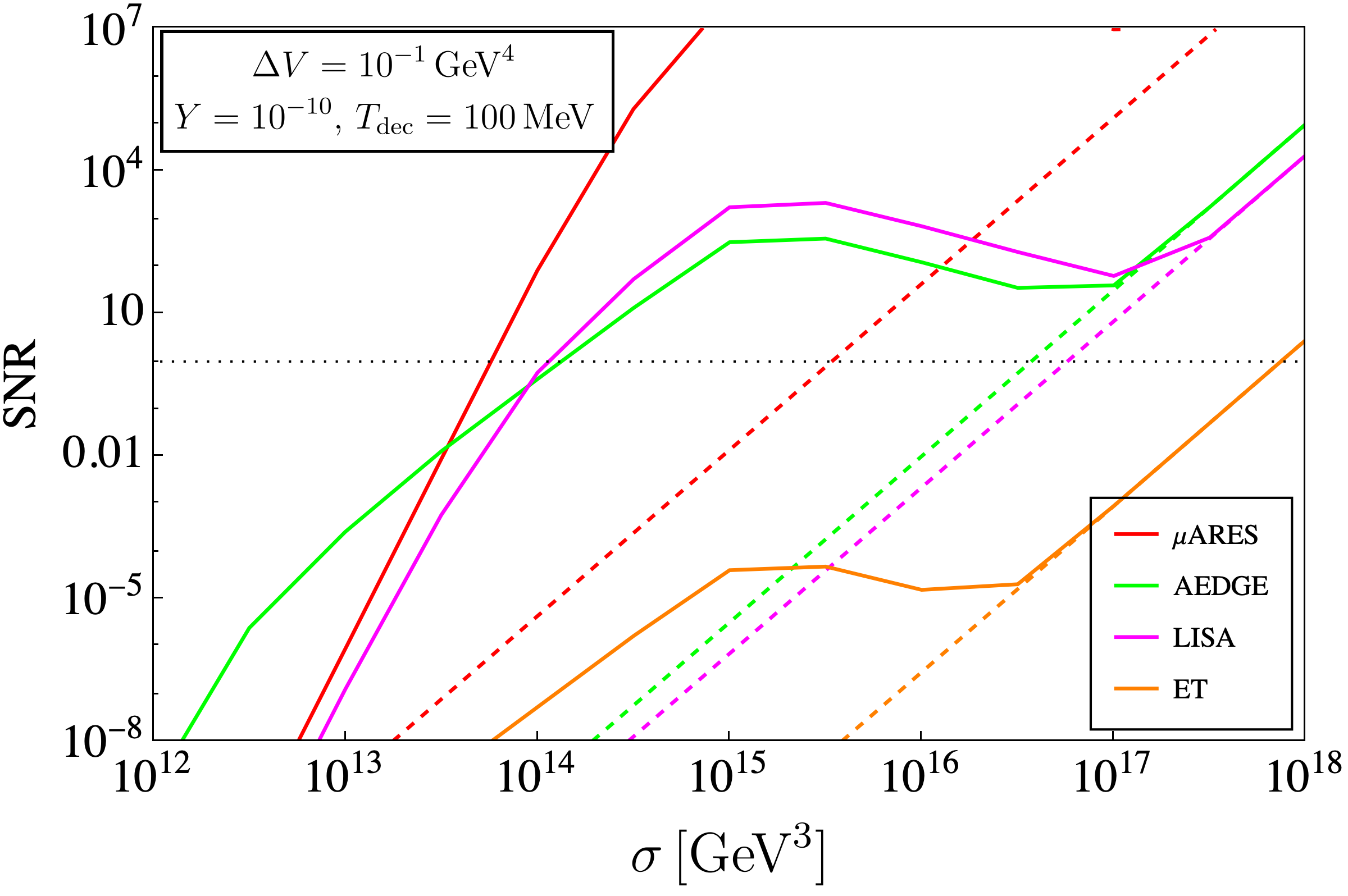}
 \caption{\it
 Plot of SNR for GW from conventional DW network (dashed) and DW network+spherons (solid).
 The horizontal dotted line indicates $\mr{SNR}=1$.
 }
 \label{fig:SNR-plt}
\end{figure}

We now wish to present the calculation of the SNR. A similar method to what we describe in this section, up to some small updates, has also been used in~\cite{Baldes:2021aph}. For details regarding the SNR estimations see e.g.~\cite{Allen:1997ad,Kudoh:2005as,Thrane:2013oya,Caprini:2019pxz,Brzeminski:2022haa}. Interferometers measure displacements in terms of a so-called dimensionless strain-noise $h_\text{GW}(f)$ that is related to the GW amplitude and can be converted into the corresponding  energy density \cite{Garcia-Bellido:2021zgu}
\begin{align}
    h^2 \Omega_\text{exp}(f)  = \frac{2\pi^2 f^2}{3 H_0^2} h_\text{GW}(f)^2 h^2,
\end{align}
with $H_0 = h\times 100 \;\text{(km/s)}/\text{Mpc}$ being the Hubble rate today. 
We compute the SNR for a given or projected experimental noise sensitivity curve $\Omega_\text{exp}(f)h^2$ in order to assess the detection probability of the GW background via the following prescription~\cite{Thrane:2013oya,Caprini:2015zlo}
\begin{align}
     \text{SNR}\equiv \sqrt{2t_\mr{obs} \int_{f_\text{min}}^{f_\text{max}} \text{d}f \left(\frac{ h^2 \Omega_\text{GW}(f)}{h^2 \Omega_\text{exp}(f)}\right)^2 }   \,, \label{eq:SNR}
\end{align}
where $h=0.7$ and  
the observation period $t_\mr{obs}$ is taken to be $20$ years for Gaia, THEIA and SKA, and $4$ years for the others.

One may consider the effective dark radiation bounds during Big Bang Nucleaosysthesis (BBN) and CMB decoupling. In particular,
the energy density of the stochastic GW background needs to be smaller than the limit on dark radiation which is depicted in $\Delta N_\text{eff}$ from BBN and CMB observations. This is because the gravitons behave as massless relativistic degrees of freedom. Any change of the number of effective
relativistic degrees of freedom ($N_\text{eff}$) at the time of recombination is usually set by the relation~\cite{Maggiore:1999vm}
\begin{align}
    \int_{f_\text{min}}^{\infty} \frac{\text{d}f}{f}  h^2 \Omega_\text{GW}(f)  \leq 5.6\times10^{-6}\;\Delta N_\text{eff} \,. \label{eq:darkrad}
\end{align}
While the lower limit for the integration is $f_\text{min}\simeq 10^{-10}\, \text{Hz}$ for BBN and $f_\text{min}\simeq 10^{-18}\, \text{Hz}$ for the CMB bounds, in practice, when, e.g., several GW spectra are depicted simultaneously, as a first-order estimate, one uses the approximation to ignore the frequency dependence and to set bounds just on the energy density of the peak for a given GW spectrum; this is shown as 
\begin{align}
    h^2 \Omega_\text{GW}^\text{Peak} \leq   5.6\times10^{-6}\;\Delta N_\text{eff} \,. \label{eq:darkrad2}
\end{align}
We consider the constraints on $\Delta N_{\rm eff}$ from BBN and the PLANCK 2018 limits~\cite{Planck:2018vyg}, as well as future reaches of CMB experiments such as CMB-S4~\cite{CMB-S4:2020lpa, CMB-S4:2022ght} and CMB-HD~\cite{Sehgal:2019ewc,CMB-HD:2022bsz}.

In order to calculate the SNR or show sensitivity curves in Fig.~\ref{fig:GW-spectrum-sph}, we need information of noise $h_\mr{GW}$ or $\Omega_\mr{exp}$.
Let us summarize the references from which we read off the information of noise. 
\begin{itemize}
\item Gaia and THEIA: noise energy density $\Omega_\mr{exp}$ from Ref.~\cite{Garcia-Bellido:2021zgu}
\item LISA: noise energy density $\Omega_\mr{exp}$ from Ref.~\cite{LISACosmologyWorkingGroup:2022jok}
\item $\mu$ARES: noise energy density $\Omega_\mr{exp}$ from Ref.~\cite{Sesana:2019vho}
\item DECIGO: strain-noise $h_\mr{GW}$  from Ref.~\cite{Kawamura:2020pcg}
\item AEDGE: strain-noise $h_\mr{GW}$ from Ref.~\cite{AEDGE:2019nxb}
\item BBO: strain-noise $h_\mr{GW}$ from Refs.~\cite{Corbin:2005ny,Harry:2006fi} 
\item ET: strain-noise $h_\mr{GW}$ from Ref.~\cite{Hild:2010id}
\item CE: strain-noise $h_\mr{GW}$ available on https://dcc.cosmicexplorer.org/CE-T2000017/public
\item SKA 20 years: power-law integrated curve directly taken from Ref.~\cite{Breitbach:2018ddu} with an appropriate factor to compensate for the difference of the threshold SNR.
\end{itemize}


Figure~\ref{fig:SNR-plt} shows plots of SNR for several future GW experiments to compare conventional DW network (dashed) and DW network+spherons (solid). 
The horizontal dotted line indicates $\mr{SNR}=1$.
Clearly one can see the enhancement of SNR thanks to the spherons.

\bibliographystyle{jhep}
\bibliography{Bibliography}


\end{document}